\documentclass[aps,prb,showpacs,twocolumn]{revtex4-1}
\usepackage{amssymb}
\usepackage{amsmath}
\usepackage{graphicx}
\usepackage{epsfig}

\begin{document}

\title{Quantum phase transition without gap opening}
\author{X. M. Yang${}^1$, G. Zhang${}^2$ and Z. Song${}^1$}
\email{songtc@nankai.edu.cn}
\affiliation{${}^1$School of Physics, Nankai University, Tianjin 300071, China \\
${}^2$College of Physics and Materials Science, Tianjin Normal University,
Tianjin 300387, China}

\begin{abstract}
Quantum phase transitions (QPTs), including symmetry breaking and
topological types, always associated with gap closing and opening. We
analyze the topological features of the quantum phase boundary of the XY
model in a transverse magnetic field. Based on the results from graphs in
the auxiliary space, we find that gapless ground states at boundary have
different topological characters. On the other hand, In the framework of
Majorana representation, the Majorana lattice is shown to be two coupled SSH
chains. The analysis of the quantum fidelity for the Majorana eigen vector
indicates the signature of QPT for the gapless state. Furthermore numerical
computation shows that the transition between two types of gapless phases
associates with divergence of second-order derivative of groundstate energy
density, which obeys scaling behavior. It indicates that a continues QPT can
occur among gapless phases. The underlying mechanism of the gapless QPT is
also discussed. The gap closing and opening are not necessary for a QPT.
\end{abstract}

\maketitle



\section{Introduction}

\label{sec_intro}

Understanding phase transitions is one of challenging tasks in condensed
matter physics. No matter which types of a quantum phase transition (QPT),
the groundstate wave function undergoes qualitative changes \cite{S. Sachdev}%
. There are various signature\ phenomena manifest the critial points, such
as symmetry breaking, switch of topological invariant, divergence of
entanglement, etc.. Among them, energy gap closing and opening seem never
absent. It is important for a deeper understanding of QPTs to find out the
role of the energy gap takes during the transition. Usually, a continuous
QPT is characterized by a divergence in the second derivative of the
groundstate energy density, assuming that the first derivative is
discontinuous \cite{Vojta,Suzuki,Dutta2}. A natural question is whether\ the
gap closing and opening are necessary for a QPT.\ The aim of this paper is
to clarify the relation between gap and a second-order QPT through a
concrete system. We take a 1D quantum $XY$ model with a transverse field,
which can be mapped onto the system of spinless fermions with $p$-wave
superconductivity. It plays an important role both in\ traditional and
symmetry-protected topological QPTs, received intense study in many aspects
\cite{G. Chen,X. X. Yi,Zanardi,Chenshu1,Chenshu2,Tao Liu,Ka-Di
Zhu,WXG,Fulibin1,Fulibin2,Fulibin3,Ling-Bao Kong,Carollo,Zhu1,Zhu2} .

The quantum phase boundary of the model is well known based on the exact
solution. However,\ it mainly arises from the condition of zero energy gap.
There are many other signatures to identify the critical point, such as the
quantities of the ground state from the point of view of quantum information
theory \cite{Osterloh}, the ground-state fidelity susceptibility \cite%
{Zanardi,Chandra,De Grandi,Chen,Tribedi,Rams,Cardy,Abanin}. Then the ground
states at different locations of the boundary may belong to different
quantum phases, although they are not protected by a finite energy gap.%
\textbf{\ }In this paper, we are interested in the possible QPT\ along the
boundary, at which the ground state is always gapless state. Our approach
consists of three steps: Bogoliubov energy band, Majorana fidelity, and
finite-size scaling. Firstly, we use Bogoliubov energy band to construct a
group of graphs that can capture the charaters of quantum phases in every
regions, including all the boundaries, which indicate the distinctions of
boundaries. Secondly, we employ the fidelity of Majorana eigen vector to
detect the QPT between two gapless phases. Thirdly, we investigate the
scaling behavior of the critical region to show a gapless QPT has the same
performance as a standard continuous QPT. The result indicates that a
continues QPT can occur among gapless phases. The gap closing and opening
are not necessary for a QPT.

This paper is organized as follows. In Section \ref{sec_model}, we present
the model Hamiltonian and the quantum phase diagram. In Section \ref%
{Topological invariants}, we investigate the phase diagram based on the
geometric properties. Section \ref{Majorana ladder} gives the connection
between the model to a simpler lattice model by Majorana transformation.\ In
Section \ref{Scaling behavior}, we present the scaling behavior about the
groundstate energy density to demonstrate the characteristic of continuous
QPT among the gapless phases. Finally, we give a summary and discussion in
Section \ref{sec_summary}.

\section{Model and phase diagram}

\label{sec_model}We consider a 1D spin-$1/2$ $XY$ model in a transverse
magnetic field $\lambda $ on $N$-site lattice. The Hamiltonian has the form

\begin{equation}
H=\sum\limits_{j=1}^{N}\left( \frac{1+\gamma }{2}\sigma _{j}^{x}\sigma
_{j+1}^{x}+\frac{1-\gamma }{2}\sigma _{j}^{y}\sigma _{j+1}^{y}+\lambda
\sigma _{j}^{z}\right) ,  \label{H}
\end{equation}%
where $\sigma _{j}^{\alpha }$ ($\alpha =\pm ,$ $z$) are the Pauli operators
on site $j$, and satisfy the periodic boundary condition $\sigma
_{j}^{\alpha }\equiv \sigma _{j+N}^{\alpha }$.

Now we consider the solution of the Hamiltonian of Eq. (\ref{H}). We start
by taking the Jordan-Wigner transformation \cite{P. Jordan}%
\begin{eqnarray}
\sigma _{j}^{x} &=&-\prod\limits_{i=1}^{j-1}\left( 1-2c_{i}^{\dagger
}c_{i}\right) \left( c_{j}^{\dagger }+c_{j}\right)  \notag \\
\sigma _{j}^{y} &=&-i\prod\limits_{i=1}^{j-1}\left( 1-2c_{i}^{\dagger
}c_{i}\right) \left( c_{j}^{\dagger }-c_{j}\right)  \notag \\
\sigma _{j}^{z} &=&1-2c_{j}^{\dagger }c_{j}
\end{eqnarray}%
to replace the Pauli operators by the fermionic operators $c_{j}$. The
parity of the number of fermions

\begin{equation}
\Pi =\prod_{l=1}^{N}\left( \sigma _{l}^{z}\right) =\left( -1\right) ^{N_{p}}
\end{equation}%
is a conservative quantity, i.e., $\left[ H,\Pi \right] =0$, where $%
N_{p}=\sum_{j=1}^{N}c_{j}^{\dag }c_{j}$. Then the Hamiltonian (\ref{H}) can
be rewritten as

\begin{equation}
H=\sum_{\eta =+,-}P_{\eta }H_{\eta }P_{\eta },
\end{equation}%
where
\begin{equation}
P_{\eta }=\frac{1}{2}\left( 1+\eta \Pi \right)
\end{equation}%
is the projector on the subspaces with even ($\eta =+$) and odd ($\eta =-$) $%
N_{p}$. The Hamiltonian in each invariant subspaces has the form

\begin{figure}[tbp]
\includegraphics[ bb=63 274 490 690, width=0.45\textwidth, clip]{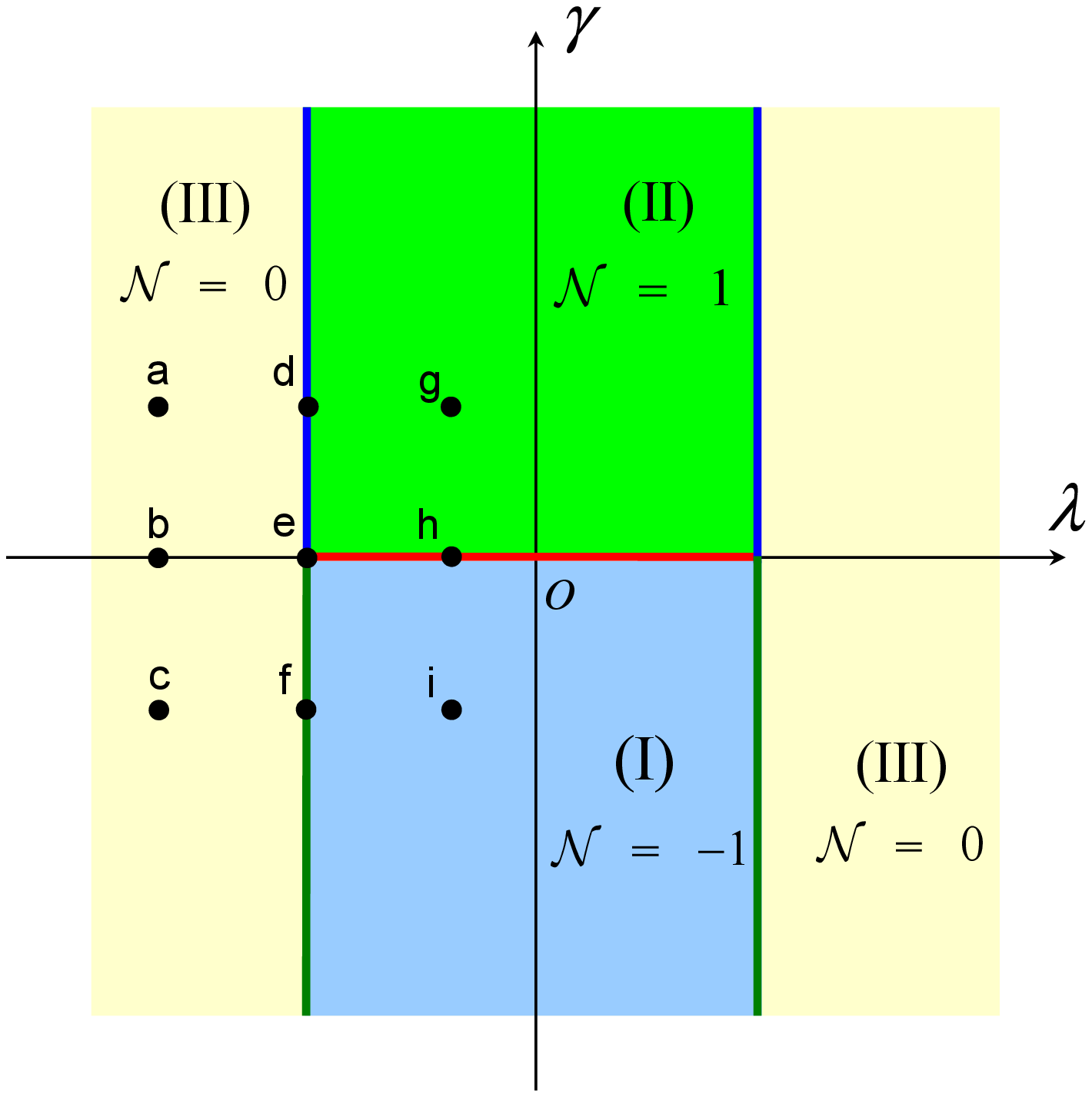}
\caption{(Color online) Phase diagram of the $XY$ spin chain on the
parameter $\protect\lambda -\protect\gamma $ plane. The blue lines indicate
the boundary, which separate the phases with winding number $0$ (yellow) and
non-trivial topological phase with winding number $1$ (green). The green
lines separate the phases with winding number $0$ (yellow) and non-trivial
topological phase with winding number $-1$ (blue). The red line separate two
phases with winding number $1$ (green)and $-1$ (blue). Several points $%
(a)-(i)$ at typical positions are indicated. The corresponding loops about
the vector $B$ are given in Fig. \protect\ref{fig2}.}
\label{fig1}
\end{figure}

\begin{eqnarray}
H_{\eta } &=&\sum\limits_{j=1}^{N-1}(c_{j}^{\dag }c_{j+1}+\gamma c_{j}^{\dag
}c_{j+1}^{\dag })-\eta (c_{N}^{\dag }c_{1}+\gamma c_{N}^{\dag }c_{1}^{\dag })
\notag \\
&&+\mathrm{H.c.}-2\lambda \sum\limits_{j=1}^{N}c_{j}^{\dag }c_{j}+N\lambda .
\end{eqnarray}%
Taking the Fourier transformation

\begin{equation}
c_{j}=\frac{1}{\sqrt{N}}\sum\limits_{k_{\pm }}e^{ik_{\pm }j}c_{k_{\pm }},
\end{equation}%
for the Hamiltonians $H_{\pm }$, we have

\begin{eqnarray}
H_{\eta } &=&-\sum\limits_{k_{\eta }}[2\left( \lambda -\cos k_{\eta }\right)
c_{k_{\eta }}^{\dag }c_{k_{\eta }}  \notag \\
&&+i\gamma \sin k_{\eta }(c_{-k_{\eta }}c_{k_{\eta }}+c_{-k_{\eta }}^{\dag
}c_{k_{\eta }}^{\dag })-\lambda ],  \label{Kiteav model}
\end{eqnarray}%
where the momenta $k_{+}=2\left( m+1/2\right) \pi /N$, $k_{-}=2m\pi /N$, $%
m=0,1,2,...,N-1$. Employing the Bogoliubov transformation%
\begin{equation*}
\gamma _{k_{\eta }}=\cos \theta _{k_{\eta }}c_{k_{\eta }}^{\dagger }+i\sin
\theta _{k_{\eta }}c_{-k_{\eta }},
\end{equation*}%
where%
\begin{equation}
\tan \left( 2\theta _{k_{\eta }}\right) =\frac{\gamma \sin k_{\eta }}{%
\lambda -\cos k_{\eta }},
\end{equation}%
one can recast Hamiltonian $H_{\eta }$ to the diagonal form%
\begin{equation}
H_{\eta }=\sum\limits_{k_{\eta }}\epsilon _{k_{\eta }}(\gamma _{k_{\eta
}}^{\dag }\gamma _{k_{\eta }}-\frac{1}{2}),  \label{H_+/-}
\end{equation}%
with spectrum being%
\begin{equation}
\epsilon _{k_{\eta }}=2\sqrt{\left( \lambda -\cos k_{\eta }\right)
^{2}+\gamma ^{2}\sin ^{2}k_{\eta }}.  \label{spectrum}
\end{equation}%
The lowest energy in $\eta $ subspace is $-\frac{1}{2}\sum_{k_{\eta
}}\epsilon _{k_{\eta }}$ for $\lambda <1$, while $-\frac{1}{2}\sum_{k_{\eta
}}\epsilon _{k_{\eta }}+\frac{1-\eta 1}{2}\epsilon _{0}$ for $\lambda >1$.
The groundstate energy for finite $N$ is the foundation for the analysis of
scaling behavior in the Sec. \ref{Scaling behavior}. In the thermodynamical
limit, the difference between two subspaces can be neglected and the
groundstate energy density can be expressed as%
\begin{equation}
\varepsilon _{g}=-\frac{1}{4\pi }\int_{-\pi }^{\pi }\epsilon _{k}dk,
\end{equation}%
by taking $k=k_{\eta }$. The quantum phase boundary can be obtained from $%
\epsilon _{k}=0$\ as%
\begin{equation}
\lambda =\pm 1,\text{and }\gamma =0\text{ for }\left\vert \lambda
\right\vert <1.  \label{boundaries}
\end{equation}

The phase diagram is presented in Fig. \ref{fig1}. There are four regions
separated by five lines as boundaries of quantum phases. In the next
section, quantum phases and boundaries will be examined from the geometrical
point of view.

\section{Topological invariants}

\label{Topological invariants}In this section, we will investigate the
topological characterization for the phase diagram. We demonstrate this
point by rewriting the Hamiltonian in the form

\begin{equation}
H=\sum_{k>0}\left(
\begin{array}{cc}
c_{k}^{\dag } & c_{-k}%
\end{array}%
\right) h_{k}\left(
\begin{array}{c}
c_{k} \\
c_{-k}^{\dag }%
\end{array}%
\right) ,
\end{equation}%
where%
\begin{equation}
h_{k}=2\left(
\begin{array}{cc}
\cos k-\lambda & i\gamma \sin k \\
-i\gamma \sin k & \lambda -\cos k%
\end{array}%
\right) .
\end{equation}%
The core matrix can be expressed as%
\begin{equation}
h_{k}=\mathbf{B}\left( k\right) \cdot \mathbf{\sigma }_{k},  \label{H_BS}
\end{equation}%
where the components of the auxiliary field $\mathbf{B}\left( k\right)
=(B_{x},B_{y},B_{z})$\ are%
\begin{equation}
\left\{
\begin{array}{l}
B_{x}=-2\gamma \sin k \\
B_{y}=2\cos k-2\lambda \\
B_{z}=0%
\end{array}%
\right. .  \label{parameter eq}
\end{equation}%
The Pauli matrices $\mathbf{\sigma }_{\mathbf{k}}$\ are taken as the form

\begin{equation}
\sigma _{x}=\left(
\begin{array}{cc}
0 & -i \\
i & 0%
\end{array}%
\right) ,\sigma _{y}=\left(
\begin{array}{cc}
1 & 0 \\
0 & -1%
\end{array}%
\right) ,\sigma _{z}=\left(
\begin{array}{cc}
0 & 1 \\
1 & 0%
\end{array}%
\right) .
\end{equation}%
The winding number of a closed curve in the auxiliary $B_{x}B_{y}$-plane
around the origin\textbf{\ }is defined as%
\begin{equation}
\mathcal{N}=\frac{1}{2\pi }\oint_{C}\mathbf{(}\hat{B}_{y}\mathrm{d}\hat{B}%
_{x}-\hat{B}_{x}\mathrm{d}\hat{B}_{y})  \label{winding N}
\end{equation}%
where the unit vector $\mathbf{\hat{B}}\left( k\right) =\mathbf{B}\left(
k\right) /\left\vert \mathbf{B}\left( k\right) \right\vert $. $\mathcal{N}$
is an integer representing the total number of times that curve travels
counterclockwise around the origin. Actually, the winding number is simply
related the loop described by equation

\begin{equation}
\frac{\left( B_{x}\right) ^{2}}{4\gamma ^{2}}+\frac{(B_{y}+2\lambda )^{2}}{4}%
\mathbf{=}1,  \label{elllipse}
\end{equation}

\begin{figure}[tbp]
\includegraphics[ bb=54 252 465 775, width=0.45\textwidth, clip]{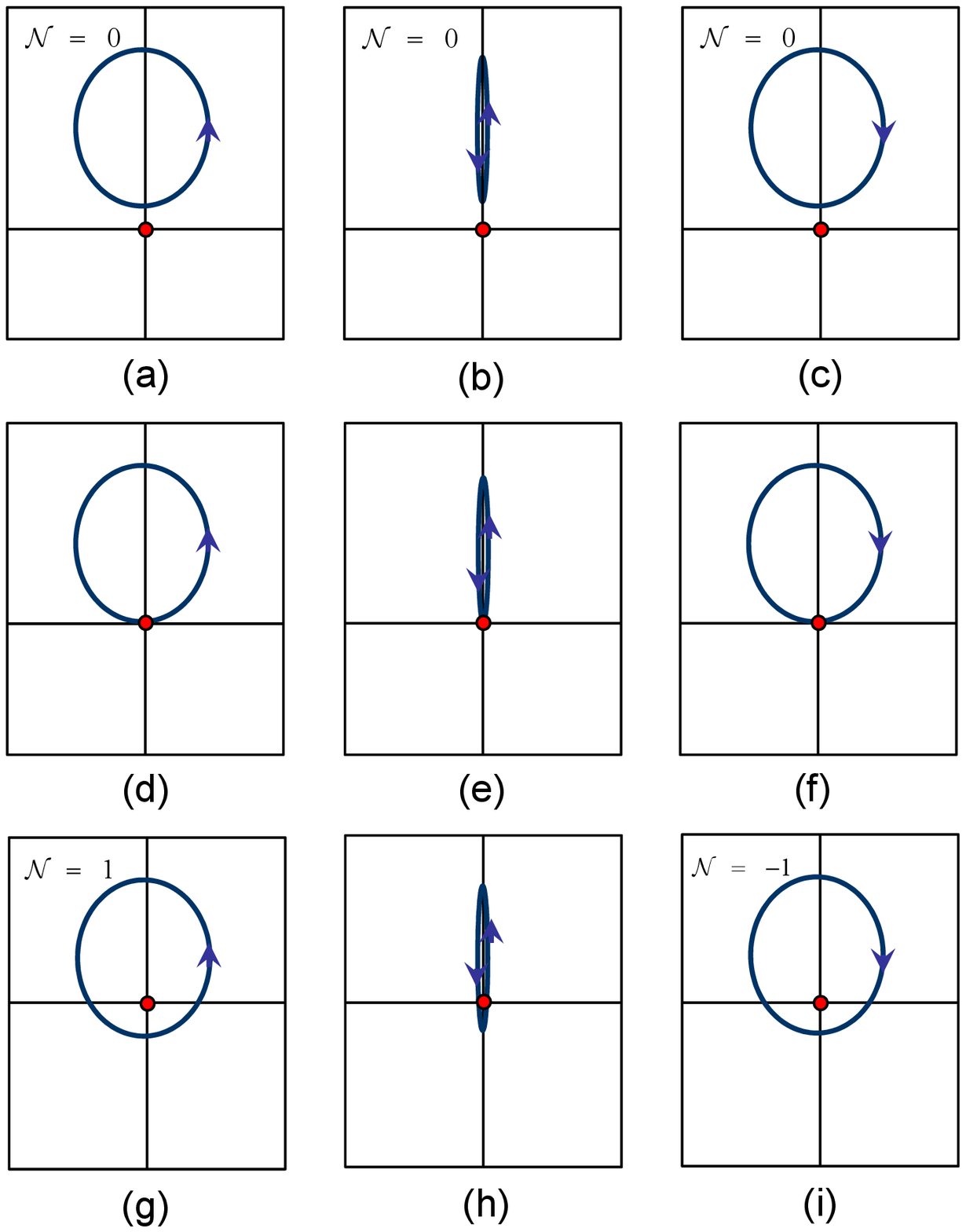}
\caption{(color online) Schematic illustration of nine types of phases by
the geometry of graphs in the auxiliary $B_{x}B_{y}$- plane. Red filled
circle indicates the origin. (a), (b) and (c) present graphs of trivial
topological phase with zero winding number. (g) and (i) present graphs of
non-trivial topological phases with winding number $\pm 1$, which are
separated by (h). (d), (e) and (f) are graphs of boundary. They correspond
to ellipses with various shapes, but passing through the origin. We see that
from (d) to (f), the graph becomes a segment. Although (d) and (f) have the
same shape, they have opposite directions, indicating two different types of
gapless phases.}
\label{fig2}
\end{figure}
which presents a normal ellipse in the $B_{x}B_{y}$-plane. The shape and
rotating direction of the ellipse dictated by the parameter equation (\ref%
{parameter eq}) have the following symmetries. First of all, taking $\gamma
\rightarrow -\gamma $, we have $[B_{x}(k),B_{y}(k)]\rightarrow \lbrack
B_{x}(-k),B_{y}(-k)]$, representing the same ellipse but with opposite
rotating direction. Secondly, taking $\lambda \rightarrow -\lambda $, we
have $[B_{x}(k),B_{y}(k)]$ $\rightarrow \lbrack B_{x}(k),B_{y}(k)+4\lambda ]$%
, representing the same ellipse but with a $4\lambda $\ shift in $B_{y}$,
while the $4\lambda $\ shift cannot affect the relation between the graph
and the origin. We plot several graphs at typical positions in Fig. \ref%
{fig2} to demonstrate the features of different phases from the geometric
point of view. We are interested in the loops for the parameters at the
boundaries in Eq. (\ref{boundaries}). (i) For $\lambda =\pm 1$ and $\gamma
\neq 0$, the ellipse always passes through the origin $(0,0)$ one time.
Along the boundary, only the length of the semiaxis of the ellipse changes.
(ii) For $\gamma =0$ and $\left\vert \lambda \right\vert \leqslant 1$, the
loop reduces to a segment, passing through the origin twice times. Along the
boundary, only the length of the segment varies. According to the connection
between loops and QPTs \cite{Gang}, when a loop passes through the origin,
the first derivative of groundstate energy density experiences a
non-analytical point. In general, this process is associated with a gap
closing. However, we note that when parameters vary along $\lambda =\pm 1$
and pass through $\gamma =0$, there is always one point in the curve at the
origin, while the ellipse becomes a segment. In the Appendix, we show that
such a system also experiences a non-analytical point but without associated
gap closing and opening.

\begin{figure}[tbp]
\includegraphics[ bb=40 600 505 750, width=0.45\textwidth, clip]{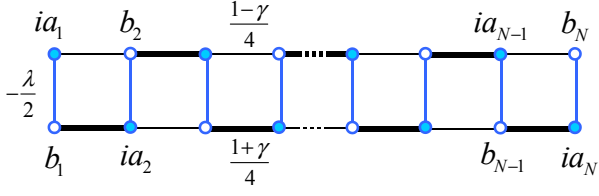}
\caption{(Color online) Lattice geometries for the Majorana Hamiltonian
described in Eq. (\protect\ref{M_matrix}), Solid (empty) circle indicates
(anti) Majorana modes. It represents two-coupled SSH chains.}
\label{fig3}
\end{figure}

\section{Majorana ladder}

\label{Majorana ladder}

In this section, we investigate the phase diagram from alternative way,
which gives a clear physical picture by connecting the obtained results to
the previous work. In contrast to last section, where a graph is extracted
from the Bogoliubov energy band, we will mark the phase diagram from the
behavior of wave functions. For quantum spin model, Majorana representation
always make things simpler since it can map a Kitaev model (like the form in
Eq. (\ref{Kiteav model})) to a lattice model in a real space with the twice
number of site of the spin system \cite{Kitaev}. The bulk-edge
correspondence has demonstrated this advantage \cite{M. Z. Hasan}.

We introduce Majorana fermion operators%
\begin{equation}
a_{j}=c_{j}^{\dagger }+c_{j},b_{j}=-i\left( c_{j}^{\dagger }-c_{j}\right) ,
\label{ab}
\end{equation}%
which satisfy the relations%
\begin{eqnarray}
\left\{ a_{j},a_{j^{\prime }}\right\} &=&2\delta _{j,j^{\prime }},\left\{
b_{j},b_{j^{\prime }}\right\} =2\delta _{j,j^{\prime }},  \notag \\
\left\{ a_{j},b_{j^{\prime }}\right\} &=&0,a_{j}^{2}=b_{j}^{2}=1.
\end{eqnarray}%
The inverse transformation is
\begin{equation}
c_{j}^{\dagger }=\frac{1}{2}\left( a_{j}+ib_{j}\right) ,c_{j}=\frac{1}{2}%
\left( a_{j}-ib_{j}\right) .
\end{equation}%
Then the Majorana representation of the Hamiltonian $H_{\eta }$ is%
\begin{eqnarray}
H_{\eta } &=&\frac{i}{4}\sum\limits_{j=1}^{N-1}\left[ (1+\gamma
)b_{j}a_{j+1}+(1-\gamma )b_{j+1}a_{j}\right]  \notag \\
&&-\frac{i\eta }{4}\left[ (1+\gamma )b_{N}a_{1}+(1-\gamma )b_{1}a_{N}\right]
\notag \\
&&+\frac{i\lambda }{2}\sum\limits_{j=1}^{N}a_{j}b_{j}+\mathrm{H.c.}.
\end{eqnarray}%
To make the structure of Majorana lattice clear, we write down the
Hamiltonian in the basis $\varphi ^{T}=(ia_{1},$ $b_{1},$ $ia_{2},$ $b_{2},$
$ia_{3},$ $b_{3},$ $...)$ and see that%
\begin{equation}
H_{\eta }=\varphi ^{T}h_{\eta }\varphi ,
\end{equation}%
where $h_{\eta }$\ represents a $2N\times 2N$ matrix. Here matrix $h_{%
\mathrm{M}}$\ is explicitly written as%
\begin{align}
h_{\eta }& =\frac{1}{4}\sum\limits_{l=1}^{N-1}[\left( 1+\gamma \right)
\left\vert l,2\right\rangle \left\langle l+1,1\right\vert +\left( 1-\gamma
\right) \left\vert l+1,2\right\rangle \left\langle l,1\right\vert ]  \notag
\\
& -\eta \frac{1}{4}[\left( 1+\gamma \right) \left\vert N,2\right\rangle
\left\langle 1,1\right\vert +\left( 1-\gamma \right) \left\vert
1,2\right\rangle \left\langle N,1\right\vert ]  \notag \\
& +\frac{\lambda }{2}\sum\limits_{l=1}^{N}\left\vert l,1\right\rangle
\left\langle l,2\right\vert +\mathrm{H.c..}  \label{M_matrix}
\end{align}

\begin{figure}[tbp]
\includegraphics[ bb=103 164 631 616, width=0.45\textwidth, clip]{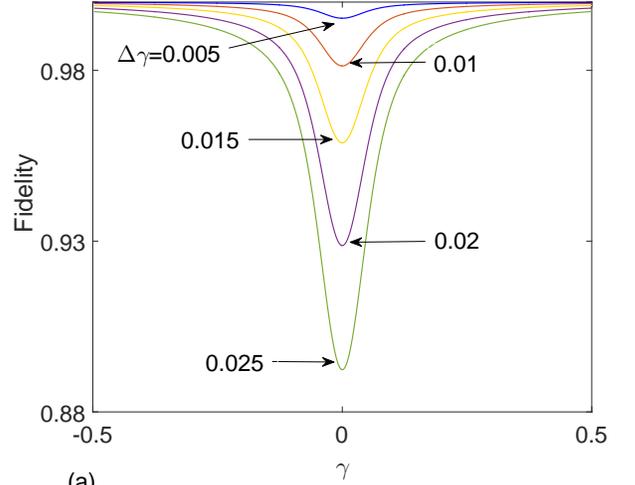} %
\includegraphics[ bb=103 164 631 616, width=0.45\textwidth, clip]{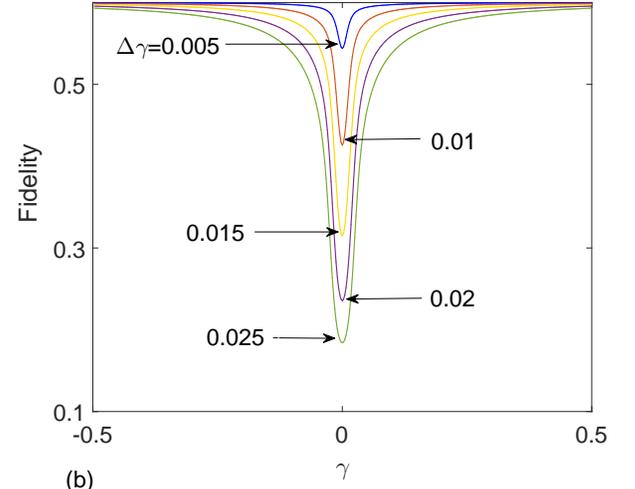}
\caption{(Color online) The fidelity of the Majorana lattice with $\protect%
\lambda =-1$ for the system with (a) $N=20$, (b) $N=100$ and various $\Delta
\protect\gamma $. }
\label{fig4}
\end{figure}

\begin{figure*}[tbp]
\includegraphics[ bb=125 154 575 579, width=0.3\textwidth,clip]{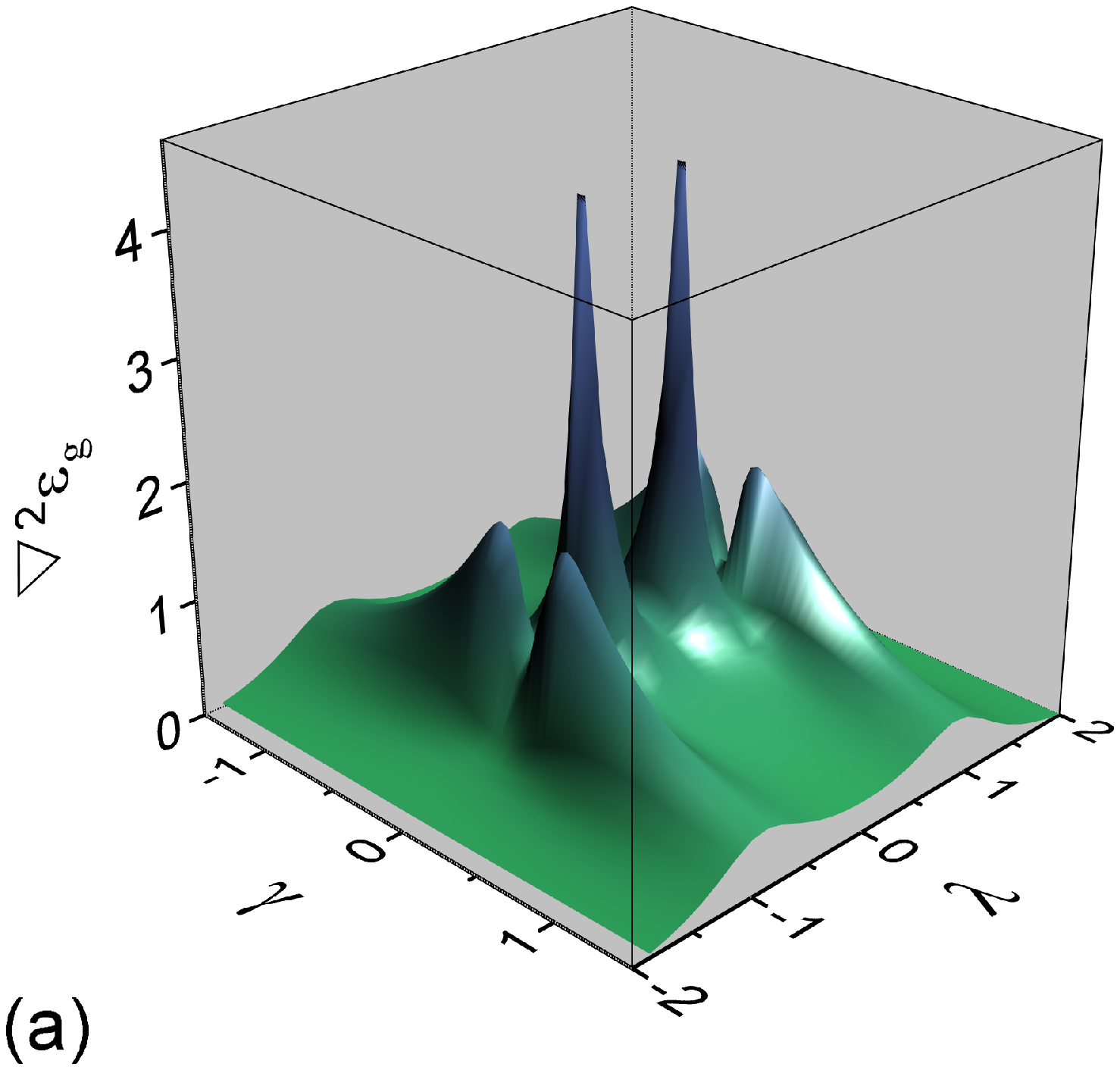} %
\includegraphics[ bb=125 152 575 579,width=0.3\textwidth, clip]{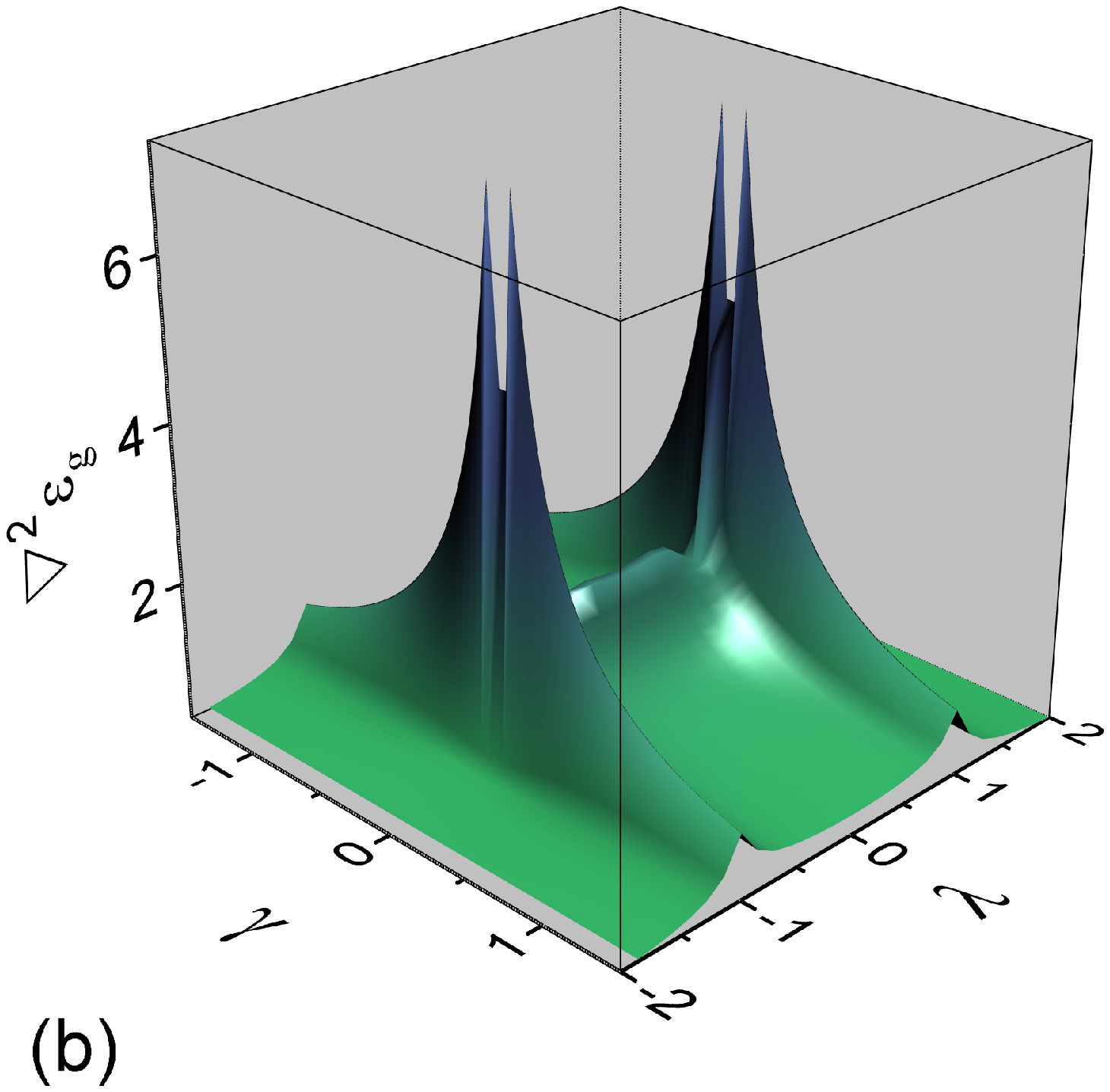} %
\includegraphics[ bb=125 152 575 579, width=0.3\textwidth, clip]{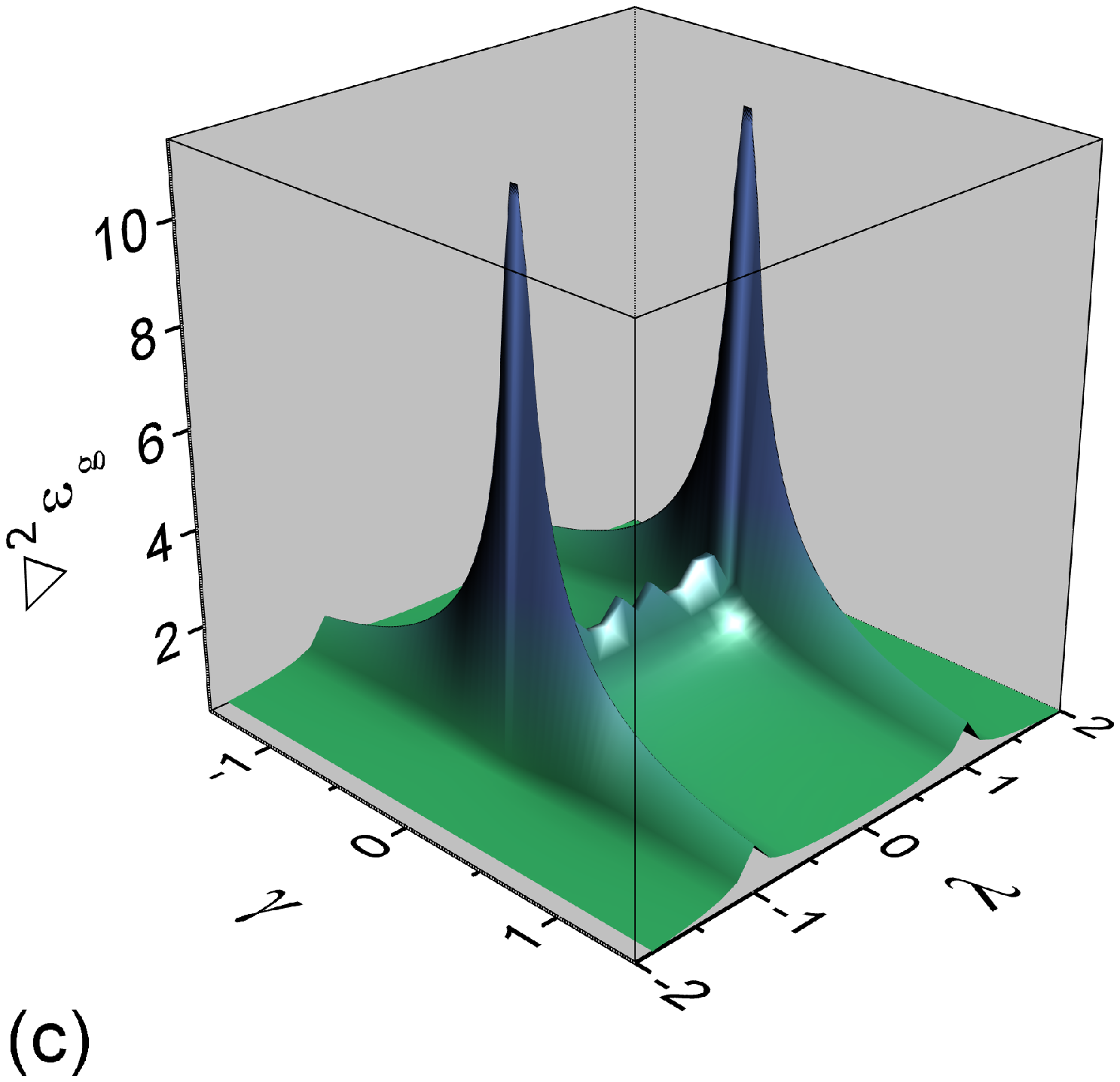}
\caption{(Color online) The Laplacian of the groundstate energy density $%
\protect\varepsilon _{g}$ as a function of the field for the case $N=8,40$
and $68$. The maxima (ridge)\textbf{\ }mark the pseudo-boundary quantum
phases. We see that there are peaks on the ridge\ near the joints for the
case of\textbf{\ }$N=68$ (see also Fig. 6(a))\textbf{.}}
\label{fig5}
\end{figure*}
where basis $\left\{ \left\vert l,\sigma \right\rangle ,\text{ }l\in \left[
1,N\right] ,\text{ }\sigma =1,2\right\} \ $is an orthonormal complete set, $%
\left\langle l,\sigma \right\vert l^{\prime },\sigma ^{\prime }\rangle
=\delta _{ll^{\prime }}\delta _{\sigma \sigma ^{\prime }}$. The basis array
is $(\left\vert 1,1\right\rangle ,$ $\left\vert 1,2\right\rangle ,$ $%
\left\vert 2,1\right\rangle ,$ $\left\vert 2,2\right\rangle ,...$ $%
\left\vert N-1,1\right\rangle ,$ $\left\vert N-1,2\right\rangle ,\left\vert
N,1\right\rangle ,\left\vert N,2\right\rangle )$, which accords with $%
\varphi ^{T}$. Schematic illustrations for structures of $h_{\eta }$\ are
described in Fig. \ref{fig3}. The structure is clearly two coupled SSH\
chains. The situation with $\eta =+$ corresponds to the case with half
quanta flux through the double SSH rings. In the previous work \cite{Li},
the gapless states in two coupled SSH chains with $\eta =-$\ has been
studied. It is shown that the quantum phase boundary $\gamma =0$ for $%
\left\vert \lambda \right\vert <1$ corresponds to topological gapless
states, while boundaries $\lambda =\pm 1$\ are trivial topological gapless
states. The joints $(\gamma ,\lambda )=(0,\pm 1)$ are boundaries separated
the gapless phases with different topological characterizations. We refer
these points as transition point between two gapless phases.

Majorana matrix in Eq. (\ref{M_matrix}) with $2N$ dimension contains all the
information of the original $H$\ in Eq. (\ref{H}) with $2^{N}$\ dimension.
Although the eigen vectors of $h_{\eta }$\ have direct relation to the
ground state of $H$, it is expected that the signature of QPT between
gapless phases can be manifested from them. We will investigate the change
of the eigen vector of $h_{+}$ along $\lambda =-1$. By the similar
procedure, the eigen problem of the equation
\begin{equation}
h_{+}\left\vert k_{+},\rho \right\rangle =\epsilon _{k_{+},\sigma
}\left\vert k_{+},\rho \right\rangle
\end{equation}%
with $\rho =\pm $, can be solved as%
\begin{equation}
\left\vert k_{+},\rho \right\rangle =\frac{1}{\sqrt{2N}}\sum_{l}(e^{ik_{+}l}%
\left\vert l,1\right\rangle +\rho e^{-i\phi }e^{ik_{+}l}\left\vert
l,2\right\rangle ),
\end{equation}%
with the eigen value%
\begin{equation}
\epsilon _{k_{+},\rho }=\frac{\rho }{2}\sqrt{\left( \cos k_{+}+1\right)
^{2}+\left( \gamma \sin k_{+}\right) ^{2}}.
\end{equation}%
Here the parameter $\phi $\ is defined as%
\begin{equation}
\tan \phi =-\gamma \tan \frac{k_{+}}{2}.
\end{equation}%
We focus on the vectors $\{\left\vert k_{+},-\right\rangle \}$\ with
negative eigen values and taking $\left\vert k_{+},-\right\rangle
=\left\vert k_{+}\right\rangle $.

We employ the quantum fidelity to detect the sudden change of the eigen
vectors, which is defined as
\begin{equation}
F\left( \gamma ,\Delta \gamma \right)
=\prod_{k_{+}}O_{k_{+}}=\prod_{k_{+}}\left\vert \langle k_{+}(\gamma -\Delta
\gamma )\left\vert k_{+}(\gamma +\Delta \gamma )\right\rangle \right\vert ,
\end{equation}%
i.e., the modulus of the overlap between two neighbor vectors with $\gamma
\pm \Delta \gamma $. Direct derivation shows that%
\begin{eqnarray}
O_{k_{+}} &\approx &1-\frac{1}{2}\left\vert \frac{\partial \left\vert
k_{+}(\gamma )\right\rangle }{\partial \gamma }\right\vert ^{2}\left( \Delta
\gamma \right) ^{2}  \notag \\
&=&1-\frac{\tan ^{2}\frac{k_{+}}{2}}{2\left( 1+\gamma ^{2}\tan ^{2}\frac{%
k_{+}}{2}\right) ^{2}}\left( \Delta \gamma \right) ^{2}.
\end{eqnarray}%
We note that the minimum of $O_{k_{+}}$\ always locates\ at $\gamma =0$ for
any values of $k_{+}$, leading to the minimum of $F\left( \gamma ,\Delta
\gamma \right) $. It indicates that the fidelity approach can be employed
for Majorana eigen vector to witness the sudden change of ground state.

\begin{figure*}[tbp]
\includegraphics[ bb=10 181 1132 545, width=1.0\textwidth, clip]{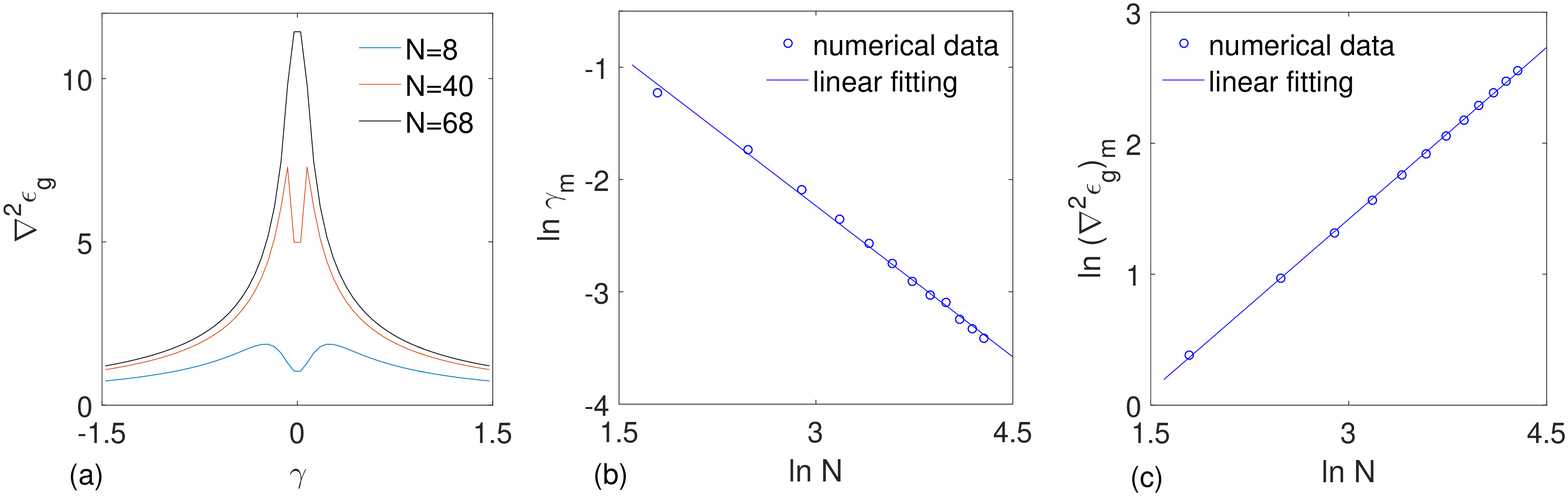}
\caption{(color online). The characteristics of second-order QPT for the
present system. (a) Plots of of $\triangledown ^{2}\protect\varepsilon _{g}$
as a function of $\protect\gamma $ for different values of $N$ at $\protect%
\lambda =1$. (b) The scaling law of pseudo critical point $\protect\gamma %
_{m}$ as a function of $N$. (c) The scaling behavior for $(\triangledown ^{2}%
\protect\varepsilon _{g})_{m}$, which is plotted as a function of $N$. The
plots are fitted by solid lines $\ln \protect\gamma _{m}=-0.897\ln N+0.456$\
and $\ln \left( \triangledown ^{2}\protect\varepsilon _{g}\right)
_{m}=0.874\ln N-1.202$.}
\label{fig6}
\end{figure*}

Fig. \ref{fig4} shows the fidelity of $h_{+}$ for a given finite system as a
function of $\gamma $ with various parameter difference $\Delta \gamma $. As
expected, the point $\gamma =0$ is clearly marked by a sudden drop of the
value of fidelity. The behavior can be ascribed to a dramatic change in the
structure of the Majorana vector, indicating the QPT at zero $\gamma $.

\section{Scaling behavior}

\label{Scaling behavior}

In this section, we investigate what happens for the groundstate energy when
a gapless QPT occurs. The non-analytical point of energy density is more
fundamental to judge the onset of a QPT. We start from the line with $%
\lambda =1$, on which the density of groundstate energy in thermodynamic
limit has the form

\begin{equation}
\varepsilon _{g}^{1}=-\frac{1}{\pi }\int_{0}^{\pi }\sqrt{(1-\cos
k)^{2}+\gamma ^{2}\sin ^{2}k}\mathrm{d}k,
\end{equation}%
where we neglect the difference between $k_{+}$\ and $k_{-}$. The first
derivative of groundstate energy density with the respect to $\gamma $\ reads%
\begin{equation}
\frac{\partial \varepsilon _{g}^{1}}{\partial \gamma }=\int_{0}^{\pi
}\digamma \left( k\right) \mathrm{d}k
\end{equation}%
where the integrand is defined as%
\begin{equation}
\mathbf{\digamma }\left( k\right) \mathbf{=-}\frac{\gamma \sin ^{2}k}{\pi
\sqrt{(1-\cos k)^{2}+\gamma ^{2}\sin ^{2}k}}.
\end{equation}%
We are interested in the divergent behavior of $\frac{\partial
^{2}\varepsilon _{g}^{1}}{\partial \gamma ^{2}}$\ when $\gamma \sim 0$.\ We
note that the main contribution to the integral\ of $\frac{\partial \digamma
}{\partial \gamma }$\ comes from the region $k\in \left[ 0,\delta \right] $\
with $\delta \ll \pi $.\ The contribution to $\frac{\partial \varepsilon
_{g}^{1}}{\partial \gamma }$\ from this region is approximately

\begin{equation}
\int_{0}^{\delta }\digamma \left( k\right) \text{d}k\approx \frac{%
4\left\vert \gamma \right\vert \gamma }{\pi }[1-\sqrt{1+\left( \frac{\delta
}{2\gamma }\right) ^{2}}],
\end{equation}%
which predicts that the first derivative of groundstate energy density along
$\lambda =1$\ has a non-analytical point at $\gamma =0$. It is the standard
characterization\ of second order QPT. It is crucial to stress that such
phase separation does not arise from the gap closing and opening.

To characterize the behavior of groundstate energy in $2$D parameter space,
we calculate the Laplacian of $\varepsilon _{g}$
\begin{equation}
\triangledown ^{2}\varepsilon _{g}=\frac{\partial ^{2}\varepsilon _{g}}{%
\partial \lambda ^{2}}+\frac{\partial ^{2}\varepsilon _{g}}{\partial \gamma
^{2}},
\end{equation}%
which will reduce to second derivative of the groundstate energy density of
the standard transverse-field Ising model \cite{S. Sachdev}\ with respect to
the transverse field $\lambda $\ when we take $\gamma =0$.\ In Fig. \ref%
{fig5} we plot the Laplacian of $\varepsilon _{g}$ for the finite sized
systems. We observe that as $N$ increases, the regions of criticality are
clearly marked by a sudden increase of the value of $\triangledown
^{2}\varepsilon _{g}$. Remarkably, there are higher order sudden increases
around the points $(\gamma ,\lambda )=(0,\pm 1)$. As before in a
conventional QPT with gap closing and opening, for instance the QPTs at $%
\lambda =\pm 1$ along $\gamma =1$, we ascribe this type of behavior to a
dramatic change in the structure of the gapless ground state. The system
undergoes a QPT along the boundary.

In order to quantify the change of the ground state when the system crosses
the critical point, we look at the value of $\triangledown ^{2}\varepsilon
_{g}$ as a function of $(\gamma ,\lambda )$\ for finite size system. The
results for systems of different size are presented in Fig. \ref{fig6}. We
find the similar scaling behavior for such kinds of QPT, which reveals the
fact that the signature of a second order QPT must not require the gap
closing.

\begin{figure}[tbp]
\includegraphics[ bb=107 300 430 700, width=0.45\textwidth, clip]{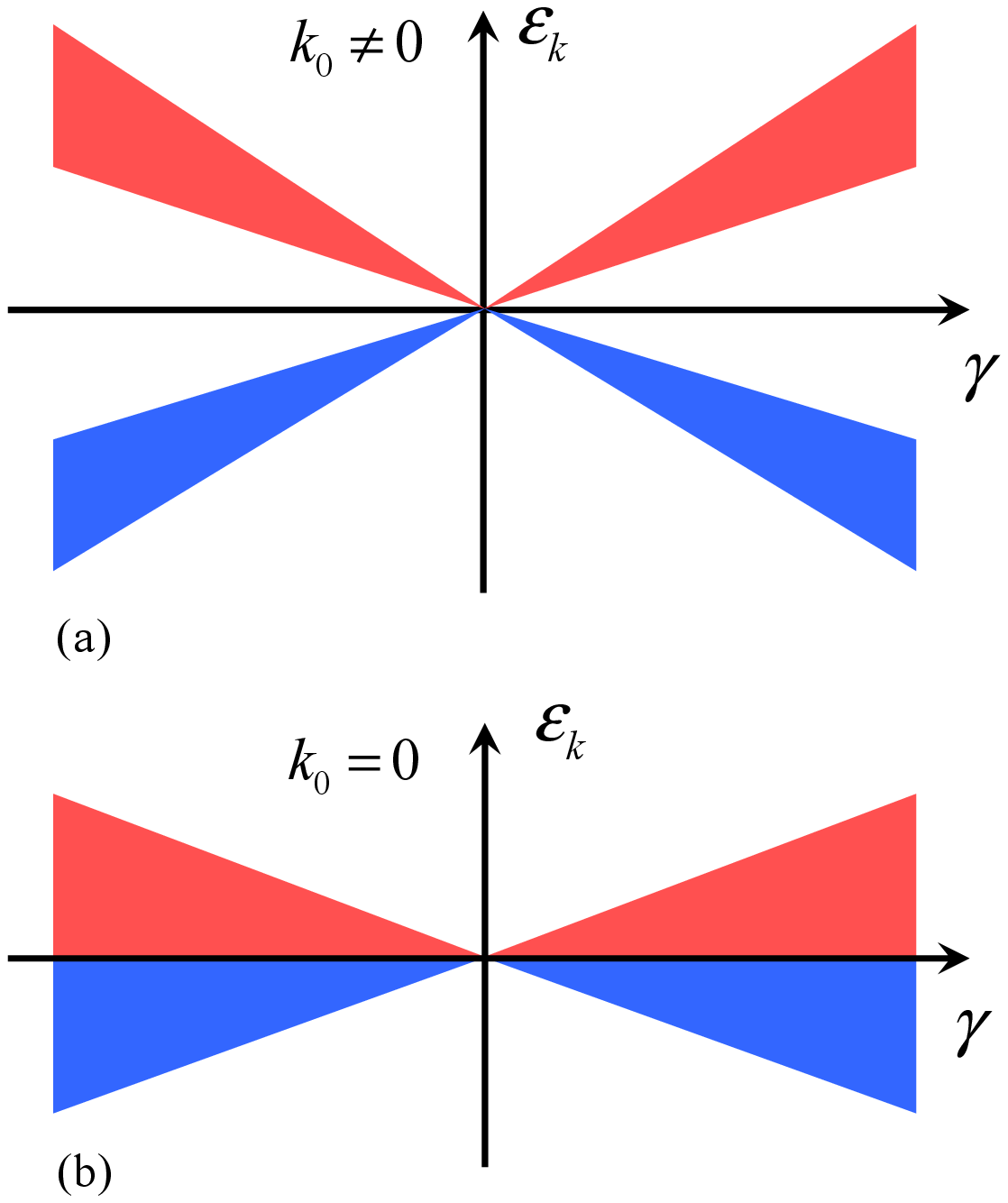}
\caption{(Color online) Energy spectra for the Hamiltonian $\sum h_{k}=%
\protect\gamma (k+k_{0})\protect\sigma _{x}$\ as a function of $\protect%
\gamma $, with (a) $k_{0}\neq 0$, and (b) $k_{0}=0$, respectively. We see
that the groundstate energies of both cases have a non-analytical points at $%
\protect\gamma =0$. However, the gap closes only at $\protect\gamma =0$\ in
the case (a).}
\label{fig8}
\end{figure}

\section{Summary and discussion}

\label{sec_summary}In summary, we have studied the the necessity of the
energy gap closing and opening for the presence of a QPT. We focused on the
joint of three types of gapless phases. The analysis based on the geometry
of the ground state and Majorana representation indicate the distinction of
the gapless phases. Numerical computation for finite size system shows that
the transitions among the gapless phases exhibit scaling behavior, which has
been regarded as the fingerprint of continuous QPT. This provides an example
to demonstrate that energy gap closing and opening is not a necessary
condition for the QPT. Mathematically, the existence of a gap depends on the
function of dispersion relation, specifically, the upper bound of the
negative band.\ The divergence of the second-order derivative of groundstate
energy density arises from the non-analytical point of the groundstate
energy, which is the summation of all negative energy levels. On the other
hand, a typical negative energy level with non-analytical point is a simple
level crossing at zero energy, for example, levels from matrix $h_{k}=\gamma
(k+k_{0})\sigma _{x}$ for $k,k_{0}\in \lbrack 0,\pi ]$. The energy gap $%
\Delta =2\gamma k_{0}$, which is zero only at $\gamma =0$, for nonzero $%
k_{0} $. However, the energy gap always vanishes for zero $k_{0}$. We plot
the spectrum for $k_{0}=0$ and $k_{0}\neq 0$, in Fig. \ref{fig8} to
illustrate this point.

\section{Appendix}

\label{sec_App}In this appendix, we demonstrate the existence of gapless QPT
through a toy model. We consider a Hamiltonian with a specific dispersion
relation $\varepsilon _{k}=\sqrt{x^{2}(k)+y^{2}(k)}$ for Bogoliubov band,
where $(x(k),y(k))$\ describes a rectangle with width $2\left\vert \gamma
\right\vert $ and length $x_{0}$\ in $xy$-plane (see Fig. 8). Here we use a
rectangle to replace a ellipse in Eq. (\ref{elllipse}) in order to simplify
the derivation. In the case of $x_{0}\gg \left\vert \gamma \right\vert $,
the main contribution to the energy is the $\varepsilon _{k}$\ at the long
sides. The parameter equations of the two long sides is
\begin{figure}[tbp]
\includegraphics[ bb=58 458 566 784, width=0.42\textwidth, clip]{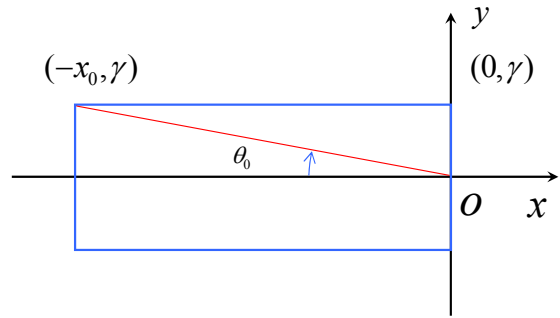}
\caption{(Color online) Schematics of a graph in auxiliary $xy$-plane for
the Bogoliubov band of a toy model, which is a rectangle. As $\protect\gamma$
varies, the origin is always at one side of the rectangle, keeping the
ground state fixed at gapless state.}
\label{app}
\end{figure}

\begin{equation}
\left\{
\begin{array}{c}
x=-\left\vert \gamma \right\vert \tan k,\text{ }(\theta _{0}\leqslant
k\leqslant \pi /2) \\
y=\pm \gamma ,%
\end{array}%
\right.
\end{equation}

where $\theta _{0}$\ is determined by tan$\theta _{0}=\left\vert \gamma
\right\vert /x_{0}$. The energy density can be expressed as%
\begin{eqnarray}
\varepsilon &=&-\frac{1}{\pi }\int_{\theta _{0}}^{\pi /2}\frac{\left\vert
\gamma \right\vert }{\sin k}\mathrm{d}k  \notag \\
&=&\frac{\left\vert \gamma \right\vert }{\pi }\ln \tan (\theta _{0}/2)=\frac{%
\left\vert \gamma \right\vert }{\pi }\ln \frac{\left\vert \gamma \right\vert
}{2x_{0}}.
\end{eqnarray}%
We note that the derivative of $\varepsilon $%
\begin{equation}
\frac{\partial \varepsilon }{\partial \gamma }=\frac{\gamma }{\pi \left\vert
\gamma \right\vert }(\ln \frac{\left\vert \gamma \right\vert }{2x_{0}}+1),
\end{equation}%
has a jump at zero $\gamma $, indicating the critical point of QPT.

\acknowledgments We acknowledge the support of the CNSF (Grant No. 11374163).

\end{document}